\documentclass[%
reprint,
superscriptaddress,
showpacs,
nofootinbib,
amsmath,
amssymb,
amsfonts,
aps,
pra,
floatfix,
showkeys
]{revtex4-1}

\RequirePackage{atbegshi}

\usepackage[utf8]{inputenc}
\usepackage[T1]{fontenc}
\usepackage[ngerman,english]{babel}
\usepackage[matha]{mathabx} 
\usepackage[dvipsnames, svgnames, table]{xcolor}
\usepackage{%
    graphicx,
    siunitx,
    multirow,
    booktabs,
    bbold,    
    pifont,   
    capt-of
}
\usepackage{subfigure}     
\renewcommand{\thesubfigure}{(\alph{subfigure})}
\makeatletter
  \renewcommand{\@thesubfigure}{\thesubfigure\space}
  \def\@currentlabel{\p@subfigure\thesubfigure}
\makeatother
\usepackage{hyperref}
\hypersetup{
    colorlinks=true,
    citecolor=black,
    linkcolor=black,
    urlcolor=black,
    anchorcolor=black,
    linktocpage
}
\usepackage{etoolbox} 
\usepackage{cleveref} 
\makeatletter
\appto{\appendix}{%
  \@ifstar{\def\theequation@prefix{A.}}%
          {}%
}
\makeatother
\crefname{figure}{Figure}{Figures}
\crefname{table}{Table}{Tables}
\crefname{equation}{Eq.}{Eqs.}
\crefname{section}{Section}{Sections}
\crefformat{appendix}{the #2Appendix#1#3} 

\definecolor{Physical_point}{RGB}{0,30,150}
\definecolor{Physical_point_text}{RGB}{0,0,150}
\definecolor{Tricritical_line}{RGB}{255,0,0}
\definecolor{Tricritical_text}{RGB}{255,0,0}
\definecolor{Z2_text}{RGB}{0,0,255}
\definecolor{Z2_line}{RGB}{0,102,255}
\definecolor{Z2_region}{RGB}{128,179,255}
\definecolor{1st_order_region}{RGB}{255,255,150}
\definecolor{1st_order_text}{RGB}{255,80,0}
\definecolor{1st_order_line}{RGB}{255,128,0}
\definecolor{O4_text}{RGB}{255,0,200}
\definecolor{O4_line}{RGB}{255,0,200}
\definecolor{O4_region}{RGB}{255,215,255}
\definecolor{1st_order_triple_region}{RGB}{170,255,204}
\definecolor{1st_order_triple_text}{RGB}{0,85,34}
\definecolor{1st_order_triple_line}{RGB}{0,180,0}
\definecolor{1st_order_quadruple_text}{RGB}{128,0,128}
\definecolor{1st_order_quadruple_line}{RGB}{170,0,212}
\definecolor{Crossover_region}{RGB}{240,240,240}
\definecolor{Crossover_text}{RGB}{100,100,100}

\graphicspath{{./figures/}}
\newcommand{\tc}[2]{\textcolor{#1}{#2}}
\newcommand{\referencename}{Ref.}
\newcommand{\referencesname}{Refs.}
\newcommand{\refcite}[1]{\referencename~\onlinecite{#1}}
\newcommand{\refscite}[2]{\referencesname~\onlinecite{#1} and~\onlinecite{#2}}

\DeclareMathOperator{\Tr}{Tr}

\newcommand{\Order}{\mathcal{O}}
\newcommand{\chidof}{\chi^2_{\text{d.o.f}}}

\newcommand{\Vector}[1]{\mathbf{#1}}
\newcommand{\Nspat}{N_\text{s}}
\newcommand{\Ntau}{N_\tau}
\newcommand{\Nc}{N_\text{c}}
\newcommand{\Nf}{N_\text{f}}
\newcommand{\hlr}{Goethe-HLR}
\newcommand{\lcsc}{\mbox{L-CSC}}
\newcommand{\Ocl}{\texttt{OpenCL}}
\newcommand{\clqcd}{CL\kern-.25em\textsuperscript{2}QCD}
\newcommand{\oqcd}{openQCD-FASTSUM}
\newcommand{\bahamas}{\texttt{BaHaMAS}}
\newcommand{\Action}{\mathcal S}
\newcommand{\ActionGauge}{\Action_{\text{g}}}
\newcommand{\ActionFermion}{\Action_{\text{f}}}

\newcommand{\Poly}{L}
\newcommand{\PolyAbs}{|{\Poly}|}
\newcommand{\Skewness}{B_3}
\newcommand{\Kurtosis}{B_4}
\newcommand{\Temp}{T}
\newcommand{\Tc}{\Temp_c}
\newcommand{\TauInt}{\tau_\text{int}}
\newcommand{\betaC}{\beta_\text{c}}
\newcommand{\ztwo}{Z_{2}}
\newcommand{\zthree}{Z_{3}}
\newcommand{\kc}{\kappa_{\ztwo}}
\newcommand{\mpi}{m_{\pi}}

\usepackage[normalem]{ulem}

\begin{document}

\title{Deconfinement critical point of lattice QCD with \texorpdfstring{$\Nf=2$}{Nf=2} Wilson fermions}

\author{Francesca Cuteri}
\email{cuteri@itp.uni-frankfurt.de}
\affiliation{
 Institut f\"{u}r Theoretische Physik, Goethe-Universit\"{a}t Frankfurt\\
 Max-von-Laue-Str.\ 1, 60438 Frankfurt am Main, Germany
}

\author{Owe Philipsen}
\email{philipsen@itp.uni-frankfurt.de}
\affiliation{
 Institut f\"{u}r Theoretische Physik, Goethe-Universit\"{a}t Frankfurt\\
 Max-von-Laue-Str.\ 1, 60438 Frankfurt am Main, Germany
}
\affiliation{
John von Neumann Institute for Computing (NIC)
GSI, Planckstr.\ 1, 64291 Darmstadt, Germany
}

\author{Alena Sch\"on}
 \email{schoen@itp.uni-frankfurt.de}
 \affiliation{
  Institut f\"{u}r Theoretische Physik, Goethe-Universit\"{a}t Frankfurt\\
 Max-von-Laue-Str.\ 1, 60438 Frankfurt am Main, Germany
}

\author{Alessandro Sciarra}
\email{sciarra@itp.uni-frankfurt.de}
\affiliation{
 Institut f\"{u}r Theoretische Physik, Goethe-Universit\"{a}t Frankfurt\\
 Max-von-Laue-Str.\ 1, 60438 Frankfurt am Main, Germany
}

\begin{abstract}
The ${\rm SU}(3)$ pure gauge theory exhibits a first-order thermal deconfinement transition due
to spontaneous breaking of its global $\zthree$ center symmetry. When heavy dynamical quarks
are added, this symmetry is broken explicitly and the transition weakens with decreasing
quark mass until it disappears at a critical point. 
We compute the critical hopping parameter 
and the associated pion mass for lattice QCD with $\Nf=2$ degenerate standard Wilson 
fermions on $\Ntau\in\{6,8,10\}$ lattices, corresponding to lattice spacings $a=\SI{0.12}{\femto\meter}$,
$a=\SI{0.09}{\femto\meter}$, $a=\SI{0.07}{\femto\meter}$,
respectively. Significant cut-off effects are observed, with the first-order region growing
as the lattice gets finer. While current lattices are still too coarse for a continuum 
extrapolation, we estimate $\mpi^c\approx \SI{4}{\giga\electronvolt}$ with a remaining systematic error of 
$\sim 20\%$. Our results allow us to assess the accuracy of
the leading-order and next-to-leading-order hopping expanded
fermion determinant used in the literature for various purposes.
We also provide a detailed investigation
of the statistics required for this type of calculation, which is useful for similar
investigations of the chiral transition.
\end{abstract}

\pacs{12.38.Gc, 05.70.Fh, 11.15.Ha}
\keywords{QCD phase diagram}
\maketitle

\section{Introduction}
\begin{figure*}[t]
    \centering
    \raisebox{-0.5\height}{\includegraphics[width=0.35\textwidth]{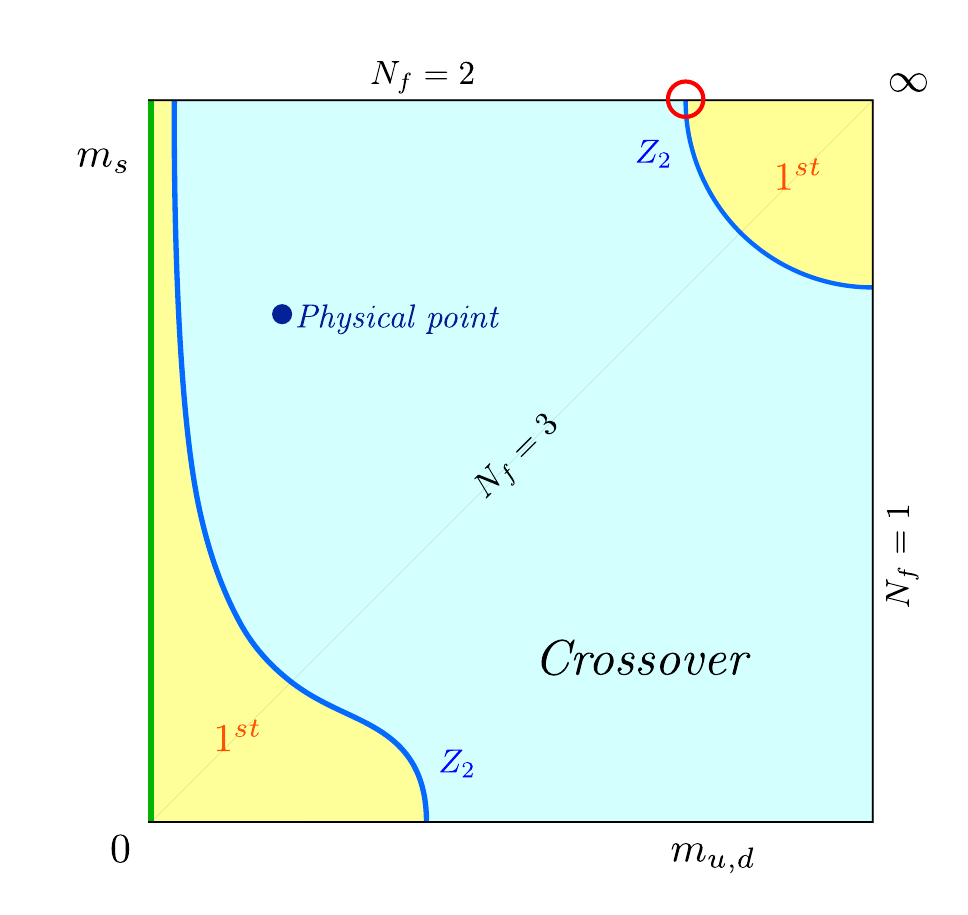}}\hfill
    \raisebox{-0.5\height}{\includegraphics[width=0.58\textwidth]{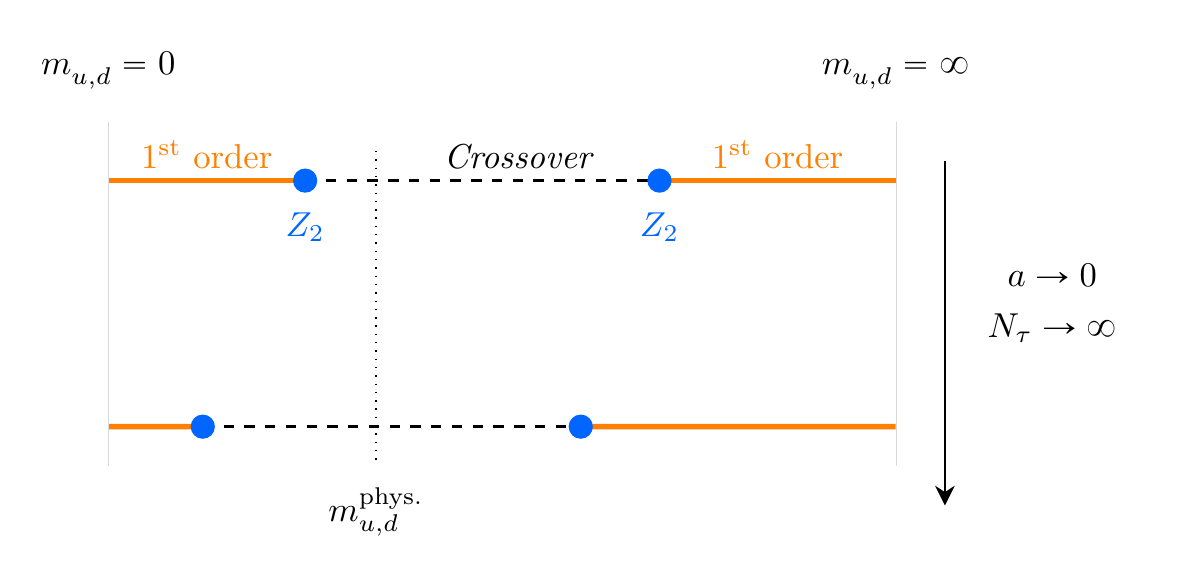}}
    \caption{%
      Left: Columbia plot for Wilson fermions on coarse lattices.
      The red circle in the heavy mass region at $\Nf=2$ indicates the critical point which we determine on $\Ntau\in\{6,8,10\}$ lattices.
      Right: Qualitative behavior of the second-order boundary point with increasing $\Ntau$, i.e., decreasing lattice spacing.
    }
    \label{fig:discretization_effects_and_columbia_plot_schematic}
\end{figure*}

For physical quark mass values, the thermal QCD transition is known to be an analytic crossover~\cite{Aoki:2006we}. 
Its continuation to small baryo-chemical potentials 
has been studied in detail 
by means of Taylor expansion or analytic continuation from imaginary chemical potential, 
with so far no hints of a non-analytic 
phase transition~\cite{Ratti:2019tvj}. Since a severe sign 
problem precludes lattice QCD simulations at finite baryon density, one way to constrain the phase diagram is
to study the thermal transition with the QCD parameters varied away from their physical values.
Non-analytic phase transitions related to the spontaneous breaking of the global center and chiral symmetries, respectively,
are explicitly seen in simulations employing unimproved fermion discretizations on coarse lattices, in
the heavy and light quark mass regime, as indicated in 
\cref{fig:discretization_effects_and_columbia_plot_schematic}. One can then study how these critical structures
evolve when a chemical potential is switched on, for an overview see~\refcite{Philipsen:2019rjq}. 

In this work we focus on the heavy mass corner, whose thermodynamics can be addressed also at finite baryon chemical 
potentials, either by means of effective lattice theories obtained from hopping expansions~\cite{Fromm:2011qi,Saito:2013vja,Aarts:2016qrv}, 
or by effective Polyakov loop theories in the continuum~\cite{Fischer:2014vxa,Lo:2014vba}. In order to assess
the accuracy of these approaches and their possible extensions to light quarks, reliable benchmarks 
are warranted.
Moreover, the phase transitions in this parameter regime are interesting 
in their own right. In the infinite quark mass limit, the theory reduces to $SU(3)$ pure gauge theory with its first-order 
transition~\cite{Boyd:1996bx},
which is caused by the spontaneous breaking of the $\zthree$ center symmetry. 
Recently, the latent heat associated with this transition has also been determined~\cite{Shirogane:2016zbf}.
When dynamical quarks are added to the theory, the
center symmetry is broken explicitly and the phase transition weakens, i.e, the latent heat decreases
until it vanishes at a critical quark mass. The value of the critical quark mass and the latent heat are thus
intimately related by the dynamics of the deconfinement transition, so that valuable non-perturbative insights
are possible in this parameter region.  

An early lattice investigation for $\Nf=1$ Wilson fermions was restricted to $\Ntau=4$ and, for large volumes, required mapping 
to an auxiliary model~\cite{Alexandrou:1998wv}.
The first systematic studies in the standard Wilson discretization~\cite{Saito:2011fs,Ejiri:2019csa} on $\Ntau\in\{4,6,8\}$ lattices 
are based on an effective potential for the order parameter determined by a histogram method~\cite{Saito:2013vja},
which is reweighted with a hopping expanded quark determinant from quenched configurations. 
This allows for a flexible investigation and interpolation between the $\Nf\in\{1,2,3\}$ cases. 
In this work 
we focus on $\Nf=2$ only, but simulate the full fermion determinant with no approximation beyond the lattice discretization.  
Preliminary results for $\Ntau\in\{6,8\}$ have been reported in~\refscite{Czaban:2016yae}{Cuteri:2017zcb}. 
Here we significantly increase statistics  
and add a third lattice spacing with a set of simulations on $\Ntau=10$ lattices. 
In agreement with earlier work, a shift of the deconfinement
critical point towards smaller bare quark masses is
observed as the lattice is made finer, \cref{fig:discretization_effects_and_columbia_plot_schematic} (right).
On the other hand, on the finer lattices we observe a quantitative deviation regarding the
location of the critical point with respect to the hopping expanded results~\cite{Ejiri:2019csa}.
It is interesting to observe that the direction of the cut-off effect in the bare parameter
space is the same for the critical deconfinement boundary in the heavy quark mass regime
as for the critical chiral boundary in the light quark mass regime with both
Wilson~\cite{Philipsen:2016hkv,Jin:2017jjp,Kuramashi:2020meg} and staggered
\cite{Karsch:2001nf,deForcrand:2003vyj,Bonati:2014kpa,Cuteri:2017gci,Cuteri:2018wci} discretizations
(in some other studies employing improved staggered fermion discretizations~\cite{Bazavov:2017xul,Ding:2019fzc},
no chiral critical line is seen, thus bounding a potential chiral first-order region).
The shift of the chiral $Z_2$ boundary implies an increase in the simulation cost while the
continuum is approached, which is particularly drastic
in the chiral transition region. 
This further motivates to attempt a continuum limit 
in the heavy quark mass regime first, where it should be more feasible. 

After devoting \cref{sec:action_and_symmetry} to the description of our lattice setup and of
the relevant symmetry at work in the infinite mass limit, we discuss our finite size scaling
analysis in \cref{sec:ana}.
Details on the simulations and the analysis strategy are provided in
\cref{sec:simulation_details}, followed by a critical appraisal of the growing statistics
requirement for decreasing lattice spacings in \cref{sec:tau_int}.
Finally, our results for the deconfinement critical point are reported in
\cref{sec:results_and_discussion} and 
conclusions are drawn in \cref{sec:conclusions}.

\section{Lattice action and center symmetry}\label{sec:action_and_symmetry}
We work with the standard Wilson gauge action
\begin{equation}\label{eq:numerical_setup_wilson_gauge_action}
	\ActionGauge = \frac{\beta}{3} \,\sum_{n} \sum_{\mu\le\nu} \,{\rm Re}\Bigl\{\Tr\bigl[ \mathbb{1}-P_{\mu\nu}(n) \bigr]\Bigr\}\;,
\end{equation}
with the lattice coupling $\beta = 2\Nc/g^{2}$, the plaquette $P_{\mu\nu}(n)$ and $n$ labeling the lattice sites. The standard Wilson 
fermion action for $\Nf$ mass-degenerate quarks is defined as
\begin{equation}\label{eq:numerical_setup_wilson_fermion_action}
    \ActionFermion = a^{4} \sum_{f=1}^{\Nf} \sum_{n_1,n_2} \bar\psi^f(n_1) \, D(n_1|n_2) \,\psi^f(n_2)
\end{equation}
with the fermion matrix
\begin{equation}\label{eq:numerical_setup_wilson_fermion_matrix}
    D(n_1|n_2) = \;\delta_{n_1,n_2} - \kappa\!\!\sum_{\mu=\pm 1}^{\pm 4}\bigl[ (\mathbb{1}-\gamma_\mu)U_{\mu}(n_1) \delta_{j+\hat\mu,n_2} \bigr] \;,
\end{equation}
where $\gamma_{-\mu}\equiv-\gamma_\mu$.
The bare fermion mass $m$ is adjusted via the hopping parameter
\begin{equation}\label{eq:hopping_parameter}
    \kappa = \frac{1}{2(am+4)}\;.
\end{equation}
The lattice coupling $\beta$ controls the lattice spacing $a(\beta)$, and temperature is defined as
\begin{equation}\label{eq:temperature}
    T=\frac{1}{a(\beta)N_\tau}\;.
\end{equation}
We do not use any improvements on the Wilson discretization to make sure the phase structure does not get modified by any 
unknown effects of improvement terms. E.g., the twisted mass formulation
introduces additional unphysical phases, which then require care to be avoided \cite{Ilgenfritz:2009ns}. Quite
generally, approaching the continuum limit in the right way requires knowledge of the lattice phase diagram in the 
bare parameter space, for which we stick to the simplest action.

The Polyakov loop $\Poly(\Vector{n})$ is defined by the product of all temporal links at space point $\Vector{n}$, 
closing through the periodic boundary,
\begin{equation}\label{eq:polyakov_loop}
    \Poly(\Vector{n}) = \frac{1}{3} \Tr_C \left[ \prod_{n_{0}=0}^{\Ntau-1} U_{0}(n_{0},\Vector{n}) \right]\;.
\end{equation}
Physically, it describes the (Euclidean) time evolution of a static quark.

At finite temperature, the periodic boundary in the temporal direction permits topologically non-trivial gauge 
transformations, which are twisted by a global center element of the group when crossing the boundary, 
\begin{equation}
    g(\tau+\Ntau,\Vector{n})=z\,g(\tau,\Vector{n})\;, \; g\in SU(3)\;,\; z=e^{i\frac{2\pi k}{3}}\mathbb{1}\;,
\end{equation}
with $k\in\{0,1,2\}$.
The pure gauge action is invariant under such transformations, $S_G[U^g]=S_G[U]$, while 
the Polyakov loop picks up the twist factor,
\begin{equation}
    \Poly^g(\Vector{n})= z^{-1} \Poly(\Vector{n})\;.
\end{equation}
In the pure gauge theory and in the thermodynamic limit, 
the expectation value $\langle\Poly\rangle$ is therefore an order parameter for center symmetry
breaking, with non-zero values signaling deconfinement~\cite{McLerran:1981pb}. 
In the presence of dynamical quarks, $S_F[U^g]\neq S_F[U]$ and the center symmetry is explicitly broken, i.e., \mbox{$\langle\Poly\rangle\neq0$} always,
with $1/m$ playing the role of the symmetry-breaking field.
Nevertheless, a rapid change accompanied by large fluctuations of $\langle\Poly\rangle$
still signals the deconfinement transition, 
which weakens with decreasing quark mass until a critical point is reached.

Being the endpoint of a first-order transition, this critical point will be in 
the same universality class as 3d liquid-gas transitions, i.e., the 3d-Ising model.
This is fully confirmed by our data. In the effective 3d-Ising Hamiltonian governing
the vicinity of the critical point, the Polyakov loop will then be the dominant
contribution to the magnetization-like variable, coupling to the symmetry-breaking 
magnetic field $1/m$. It is interesting to contrast this situation with 
the chiral transition in the light quark regime, where the Polyakov loop contributes
to the energy-like variable in the corresponding effective Hamiltonian, since
it is invariant under chiral transformations \cite{Clarke:2020htu}.

\section{Finite size scaling analysis \label{sec:ana}}

\begin{figure*}[t]
    \centering
    \includegraphics[width=0.88\textwidth, clip, trim=0 4mm 0 0 ]{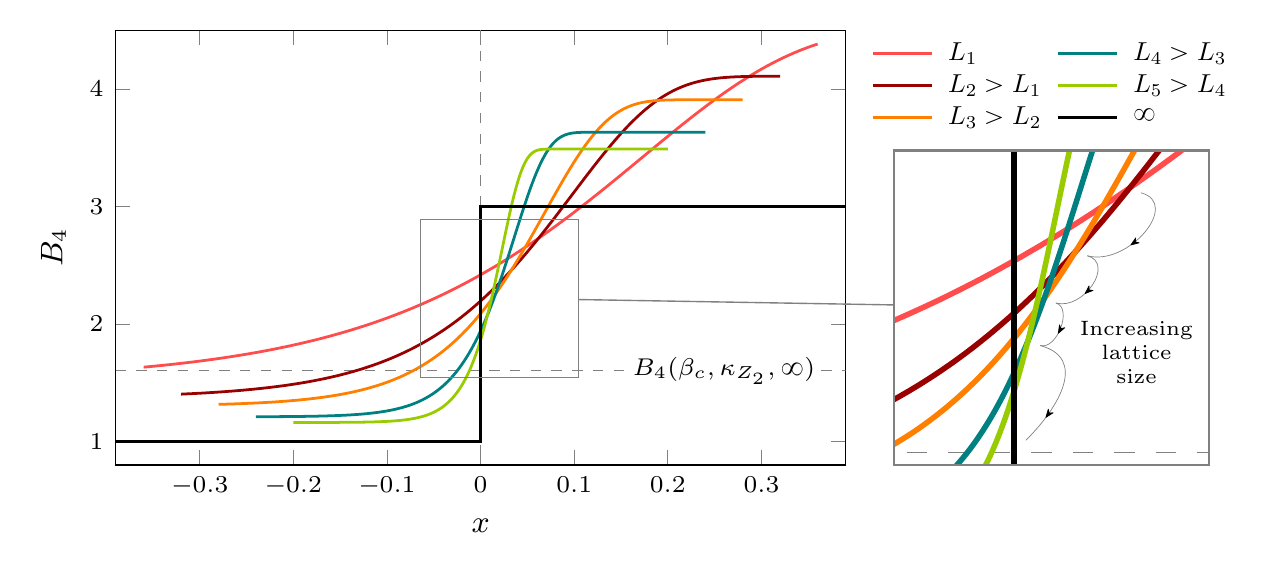}
	\caption{%
      Qualitative behavior of the kurtosis $\Kurtosis$ for spatial volumes too small to be 
      described by \cref{{eq:fitting_form_of_b4}}.
      The effect of the leading finite size correction, \cref{eq:kurtosis_finite_volume_correction_term}, is to shift 
      $\Kurtosis$ at $\kc$ (i.e. for $x=0$) to larger values and approach the universal value with growing volumes.
      The enlargement shows that different volumes do not cross in one point.
      Pairwise intersections converge to $\Kurtosis(\betaC,\kc,\infty)$ in the thermodynamic limit.
    }
    \label{fig:B4_schematic}
\end{figure*}
To determine the order of the phase transition as a function of the quark mass,
we compute standardized moments of $X\equiv\PolyAbs$, the absolute value of the Polyakov loop,
\begin{equation}\label{eq:central_moments}
    B_n(X;\,\alpha_1,\alpha_2,\ldots)=
    \frac{\bigl\langle(X-\langle X\rangle)^n\bigr\rangle}
                    {\bigl\langle(X-\langle X\rangle)^2\bigr\rangle^{\frac{n}{2}} \vphantom{\Bigl[}}\;,
\end{equation}
and study their behavior as a function of the physical volume.
Here $n\in\{3,4\}$ and $\{\alpha_n\}$ is a set of relevant physical parameters, in our case
$\beta,\kappa,\Ntau,\Nspat$ (in the following, the dependence of $B_n$ on $X$ will be
understood).

The phase boundary at a pseudocritical coupling $\betaC$ is defined by the vanishing of the skewness $\Skewness(\betaC)=0$.
We use finite size scaling of the fourth moment, the kurtosis $\Kurtosis$, trivially linked to the Binder cumulant~\cite{Binder1981},  
to locate the critical point $\kc$.
\begin{table}[t]
    \setlength{\tabcolsep}{7pt}
    \renewcommand{\arraystretch}{1.0}
    \centering
    \begin{tabular}{cccS[table-number-alignment=left]}
        \toprule
        & Crossover & $1^\text{st}$ order & {$2^\text{nd}$ order $Z_2$}\\
        \midrule
        $\Kurtosis$ & 3 & 1 & 1.604 \\
        $\nu$ & - & $1/3$ & 0.6301(4) \\
        $\gamma$ & - & 1 & 1.2372(5) \\
        $\alpha$ & - & - & 0.110(1) \\
        \bottomrule
    \end{tabular}
  \caption{Values of the kurtosis at the transition and of the relevant critical exponents~\cite{Pelissetto:2000ek}.}
  \label{tab:b4_values}
\end{table}
For $V\to\infty$, $\Kurtosis$ evaluated on the phase boundary takes the form of a step function, assuming values that are 
characteristic for the respective nature of the phase transition (c.f. \cref{tab:b4_values}).
On finite volumes, $\Kurtosis$ is an analytic curve that gradually 
approaches the step function with increasing volume. In the vicinity of the critical point $\kc$, universality implies it to be a function 
of the scaling variable $x\equiv(\kappa-\kc)\Nspat^{1/\nu}$ only, which can be expanded around $\kappa=\kc$ in a Taylor series. 
For sufficiently large volumes, the series can be truncated after the linear term, 
\begin{equation}\label{eq:fitting_form_of_b4}
    \Kurtosis(\betaC,\kappa,\Nspat) = \Kurtosis(\betaC,\kc,\infty) + a_1\, x + \Order(x^2)\;,
\end{equation}
and fitted to extract the critical parameters.

This simple picture holds for asymptotically large volumes, and
previous studies show that these can be prohibitively expensive to attain, 
cf.~\cite{Philipsen:2014rpa,Cuteri:2015qkq}.
Moreover, we find the required aspect ratios in the heavy mass region to be even
significantly larger than those in our light quark studies. The reason cannot be 
related to the lightest state in 
the spectrum, the glueball in this case, whose correlation length is shorter than that 
associated with the pion in the light quark mass regime.
Instead, a possible explanation
is that regular (non-divergent) contributions to the 
fluctuations grow 
towards the heavy mass region, so that it takes larger volumes for the diverging terms to dominate. 

It is therefore expedient to also consider the leading finite volume corrections to \cref{eq:fitting_form_of_b4}. 
Since for finite quark masses the center symmetry is explicitly broken, the Polyakov loop is no longer 
a true order parameter, but a mixture of 
the magnetization- and energy-like observables entering the effective 3d-Ising Hamiltonian which governs the vicinity 
of the critical point.
Magnetization- and energy-like observables scale with different exponents, and only for asymptotically large volumes 
the larger of them dominates to produce the simple expression \cref{eq:fitting_form_of_b4}. 
With the leading mixing correction taken into account, the kurtosis takes instead the form
\begin{align}\label{eq:kurtosis_finite_volume_correction_term}
    \Kurtosis(\betaC,\kappa,\Nspat) =&\bigl[ \Kurtosis(\betaC,\kc,\infty) + a_1 x + \Order(x^2)\bigr] \notag \\
    \times &\bigl[1+B\,\Nspat^{(\alpha-\gamma)/2\nu}\bigr],
\end{align}
where $B$ is another non-universal parameter to be extracted from a fit.
For a detailed derivation and explanation in the context of the light quark mass regime, see~\cite{Jin:2017jjp}.
As a consequence of this correction, the $\Kurtosis$ curves are no longer fitted by straight lines, and different volumes
intersect at different points, with the pairwise intersection approaching
the universal value with growing volumes. 
This is shown schematically in \cref{fig:B4_schematic}.

\section{Simulation and analysis details}\label{sec:simulation_details}
\begin{figure*}[t]
    \centering
    \includegraphics[width=0.82\textwidth, clip, trim=0 0 0 9mm]{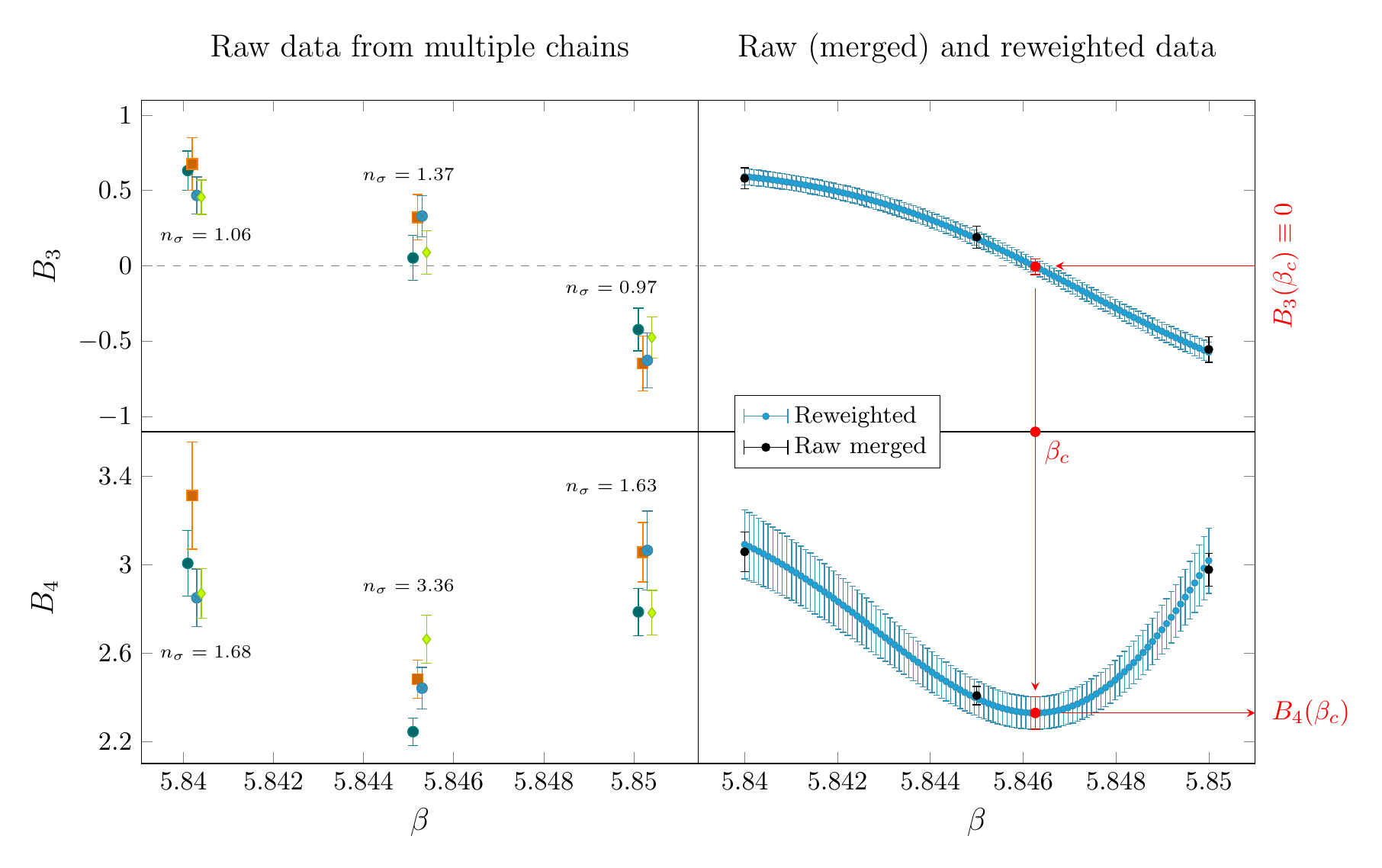}
    \caption{%
      Example of the data analysis for $\Skewness, \Kurtosis$ at $\kappa=0.1100$ on a $6\times36^3$ lattice.
      In the left plots different Markov chains are displayed,
      with data points slightly shifted horizontally for better visibility.
      The label $n_\sigma$ denotes by how many standard deviations the 
      maximally incompatible pair is apart.
      On the right, the merged raw and reweighted data for $\Skewness$ (top) and $\Kurtosis$ (bottom) are shown.
      The determination of $\betaC$ and $\Kurtosis(\betaC)$ is depicted in red.
    }
    \label{fig:B3}
\end{figure*}
We study cut-off effects by simulating three different temporal lattice extents $\Ntau\in \{6,8,10\}$, corresponding
to different lattice spacings at the respective critical couplings. 
For every value of $\Ntau$, the critical $\kc$ in the heavy quark mass region was determined
(i.e.\ the position of red circle in \cref{fig:discretization_effects_and_columbia_plot_schematic})
for five to six values of $\kappa$.
At each value of $\kappa$ up to four spatial volumes have been simulated,
corresponding to aspect ratios of $\Nspat/\Ntau\in[4,7]$.
For $\Ntau=8$, also simulations at $\Nspat=80$ have been done to give better insights into the position of $\kc$.
The deconfinement transition has been located simulating at two to four different values of $\beta$ around the pseudocritical coupling.
Configurations have been produced using a standard Hybrid Monte Carlo algorithm~\cite{Duane:1987de} with unit trajectories and the
acceptance rate tuned to stay between 75\% and 85\%. At least $5000$ thermalization trajectories have always been discarded.
Observables (the plaquette and the Polyakov loop) have then been measured on the fly for all Monte Carlo steps.
At each value of $\beta$, between 56k and 800k trajectories have been accumulated, making sure to always have over $\approx50$ independent events per $\beta$.
In total $28$, $39.5$ and $21.8$ millions trajectories have been produced for $\Ntau=6$, $\Ntau=8$ and $\Ntau=10$, respectively.
Details about simulation parameters and statistics are listed in
\cref{tab:simulation_overview_nf2_mu_zero,tab:tauint_B4,tab:tauint_B3}.

All the finite temperature simulations, except from those on the $8\times80^3$ lattices, have been performed using \texttt{v1.0} of our publicly available~\cite{CL2QCD}
\Ocl\ based \clqcd\ code~\cite{Bach:2012iw}, which is optimized to run on GPUs. The \lcsc\ supercomputer~\cite{L-CSC} at GSI in Darmstadt has been used for this set of runs.
The few simulations on aspect ratio $10$ lattices have been performed with \oqcd~\cite{openQCD-FASTSUM}, a highly MPI-optimized software that has been run on the
\hlr\ supercomputer.
In all cases, monitoring and handling of thousands of jobs has been 
efficiently automatized using the software package
\bahamas~\cite{Sciarra:2017gmt,BaHaMAS}, whose new version \texttt{v0.3.1} can also be used in conjunction with the \oqcd\ code.

In order to faster accumulate statistics and to gain better control over statistical errors, for each
parameter set we simulated four different Markov chains, except at the outermost $\beta$ values
for aspect ratio $10$ where only three Markov chains were run.
In this way a criterion to decide when the gathered statistics is large enough can be established: All replicas included in the final analysis were run until $\Skewness$ was compatible within at 
most $3$ standard deviations between all of them.
An example of one dataset can be found in \cref{fig:B3} (left).
At the beginning of the simulations, for very low statistics, this criterion can be trivially fulfilled due to very large statistical errors.
Therefore, to ensure a proper bracketing of $\betaC$ and to avoid stopping the simulations too early, it has been almost always demanded that $\Skewness$ is incompatible with zero at $1$ standard deviation at the smallest and largest $\beta$ on every chain.
Once both these guidelines are satisfied, the chains were merged to the raw data points shown in \cref{fig:B3} (right).

In order to have precision on $\betaC$, the multi-histogram method~\cite{FSReweighting}, also known as reweigthing, was used to interpolate between our measurements. 
This interpolation is repeated for the $\Kurtosis$ data, thus pinning down $\Kurtosis(\betaC)$. 
Carrying out the same procedure for each simulated value of $\kappa$, 
the resulting $\Kurtosis(\betaC,\kappa,\Nspat)$ data points can be plotted as a function of $\kappa$ and 
fitted according to \cref{eq:fitting_form_of_b4,eq:kurtosis_finite_volume_correction_term}, as it will be discussed in \cref{sec:results_and_discussion}.

Finally, to set the physical 
scale for our simulation points we use the Wilson flow parameter $\omega_{0}/a$, 
which we determined and converted using the publicly available software described in~\refcite{Borsanyi:2012zs}. 
To this end, we produced $800$ independent zero-temperature configurations on $32\times 16^{3}$ lattices for each $\kappa$ at the value of the critical coupling $\betaC$. 
As a physical quantity to parametrize the determined critical point $\kc$, we then computed the pseudoscalar meson mass $\mpi$
corresponding to those bare parameter values.
Note that for all lattice spacings considered here, $a\mpi > 1$ at and around $\kc$.
Therefore, our pion mass estimates are afflicted by large cut-off effects.
This problem naturally reduces as $\Ntau$ is increased.
All critical couplings $\betaC$, the corresponding lattice spacings $a$, $\mpi$ and  
critical temperatures $\Tc$ are summarized in \cref{tab:nf2_nt6_nt8_nt10_mu0_data} for each value of $\kappa$ and $\Ntau$.

\section{Statistics requirements towards the continuum}\label{sec:tau_int}
\begin{figure}[t]
    \centering
    \subfigure[Fixed $\Ntau.$]{\label{fig:tau_int_nt_8_B3}\includegraphics[width=0.98\columnwidth]{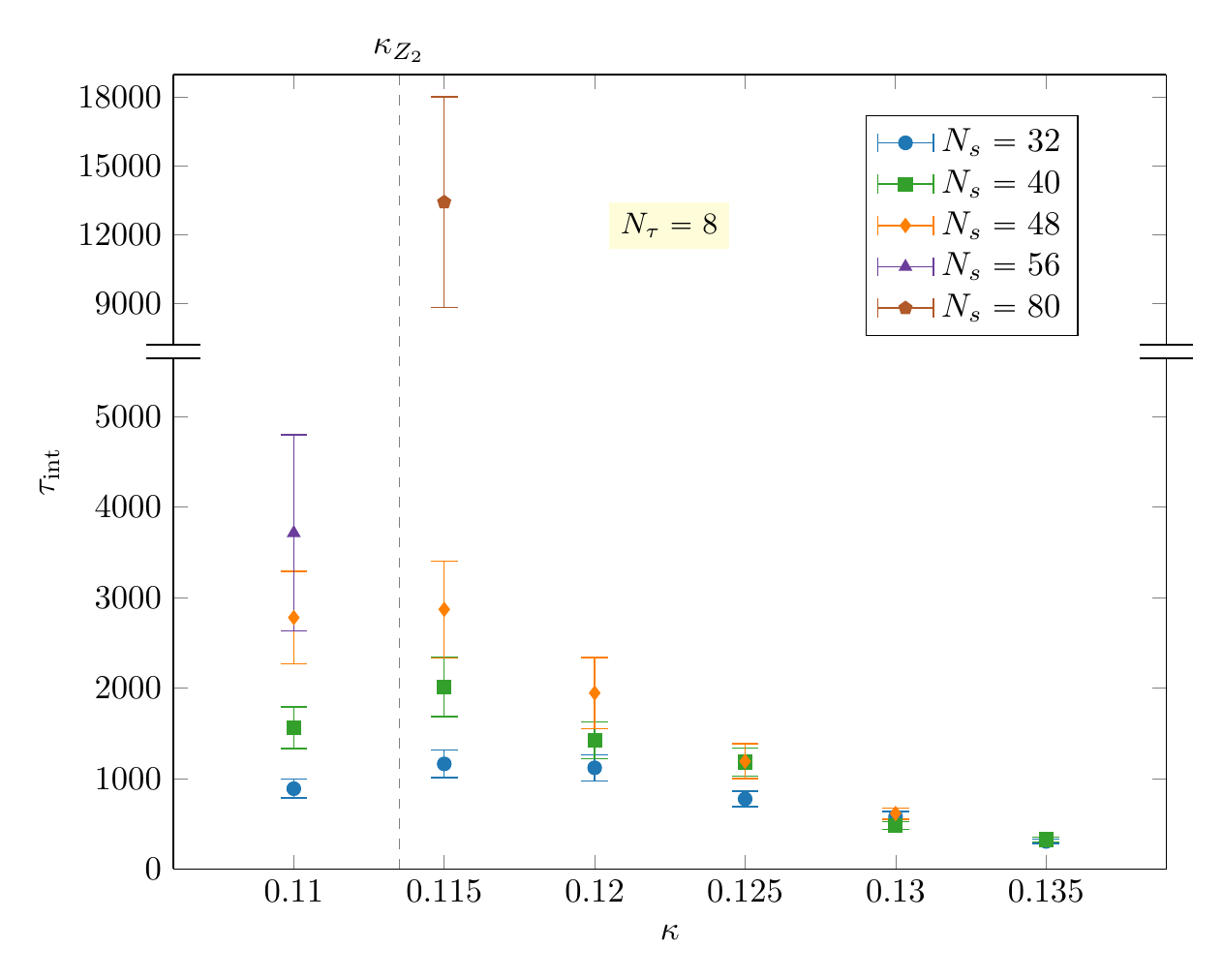}}\\
    \subfigure[Fixed aspect ratio.]{\label{fig:tau_int_aspect_ratio_6_B3}\includegraphics[width=0.96\columnwidth, clip, trim=0 0 3mm 0]{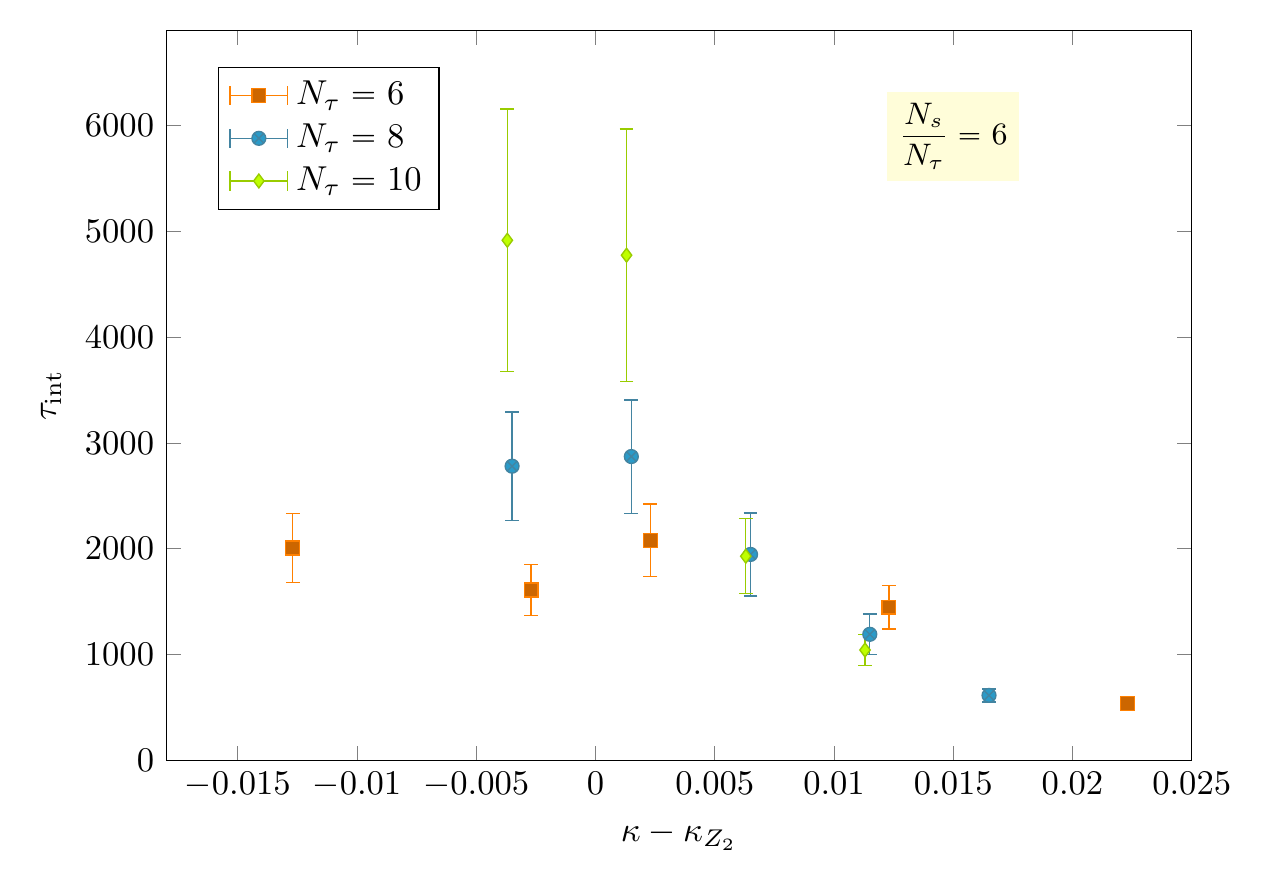}}
    \caption{%
      The integrated autocorrelation time $\TauInt$ of the skewness of the order parameter is shown for the simulated $\beta$ closest to $\betaC$ for different values of $\kappa$.
      In \subref{fig:tau_int_nt_8_B3} $\Ntau$ is kept fixed and the spatial volume is varied ($\kc$ is marked by the dashed line).
      In \subref{fig:tau_int_aspect_ratio_6_B3} $\TauInt$ is plotted against $\kappa-\kc$ for different $\Ntau$ at fixed aspect ratio.
    }
    \label{fig:tau_int_B3}
\end{figure}

A crucial parameter to judge the statistical quality of the analysed 
ensembles is the integrated autocorrelation 
time $\TauInt$ of the observables. 
Qualitatively speaking, $\TauInt$ parametrizes the memory that the simulated system has of its dynamics, which in our case
strongly depends on the order of the phase transition.
Consider a first-order transition to start with and let us briefly summarize what happens to the behaviour of the order parameter.
In a finite volume, far away from the phase transition, the system stays in a given phase.
The order parameter will fluctuate around its mean value and its probability distribution 
will be approximately Gaussian, with some associated
$\TauInt$. 
As soon as $\beta$ gets sufficiently close to $\beta_c$, the system will explore both phases and the order parameter will jump from time to time from fluctuating around its mean value in one phase to fluctuating around its mean value in the other one.
These fluctuations between the phases are slower than the fluctuations within 
one phase, since
tunneling between different phases is
exponentially suppressed by a potential barrier growing with volume \cite{Langer:1969bc}. 
As a result,
the system has a much longer-term memory since its dynamics now occurs on a larger timescale.
$\TauInt$ is related to the average tunneling rate between the two phases and thus
increases significantly.
This in turn implies that sufficiently many tunneling events in a simulation are needed to reliably estimate $\TauInt$.
At a crossover transition, instead, the distribution of the order parameter does not show a two-peak structure for any $\beta$ value, even around $\betaC$, where only its variance slightly increases.
Therefore, $\TauInt$ is expected to be relatively small in this case.
Finally, at a second-order phase transition the critical slowing down phenomenon 
will play an important role and the diverging correlation length of the system will be reflected in an increase of $\TauInt$, too.
In a finite volume this effect will be only partially felt, though, because the correlation length cannot literally diverge.

We estimate $\TauInt$ for both $\Skewness$ and $\Kurtosis$ using a \texttt{PYTHON} implementation of the
$\Gamma$-method~\cite{Wolff:2003sm}, and 
distribute our data in bins of size $2\TauInt$ to remove autocorrelations.
\cref{tab:tauint_B3,tab:tauint_B4} give an overview of $\TauInt$ for the skewness and the kurtosis (averaged over the simulated $\beta$ values)
and the number of statistically independent events obtained from the binning procedure.
The qualitative expectation outlined above is completely met in \cref{fig:tau_int_B3}, 
where for each $(\kappa,\Nspat)$ pair,
$\TauInt$ is displayed at the simulated $\beta$ closest to $\betaC$.
In \cref{fig:tau_int_nt_8_B3} the autocorrelation time is shown as 
a function of $\kappa$ and spatial volume for $\Ntau=8$.
For each volume, a maximum is seen around $\kc$, showing critical 
slowing down near a second-order transition (that maximum would be sharper 
if $\TauInt$ had been evaluated exactly at $\betaC$).
The more drastic effect, however, is the increase of $\TauInt$ with increasing volume. 
In \cref{fig:tau_int_aspect_ratio_6_B3} we compare different $\Ntau$ values at
fixed aspect ratio. As expected when approaching the continuum limit, we
observe another increase of $\TauInt$ around the critical as well as the first-order region.
The statistics for $\Ntau=10$ is effectively smaller compared to $\Ntau\in\{6,8\}$, 
which explains the larger error bars and possibly an overestimate in the first-order
region. 

Note also, that the observed rise in the autocorrelation time feeds back
into the practical organization of the simulation. 
The choice of $\beta$-values to be simulated is an optimization of having a sufficiently
narrow spacing needed for reweighting, and covering a large enough interval to 
bracket $\beta_c$, with as few values as possible. This requires monitoring and frequently
analysing running simulations in order to optimally adjust the parameters.
Since a given statistics which turned out to be sufficient on a coarser lattice, will have to be increased on a finer one,
even this process of parameter tuning requires more and more trajectories as the lattices
get finer.
Altogether, our study of the autocorrelation time illustrates the 
crucial importance and formidable difficulties of obtaining sufficient statistics
for a reliable determination of the order of the phase transition
as the continuum is approached.


\section{Results and Discussion}\label{sec:results_and_discussion}
\begin{table*}[t]
    \centering
    \includegraphics[width=\textwidth]{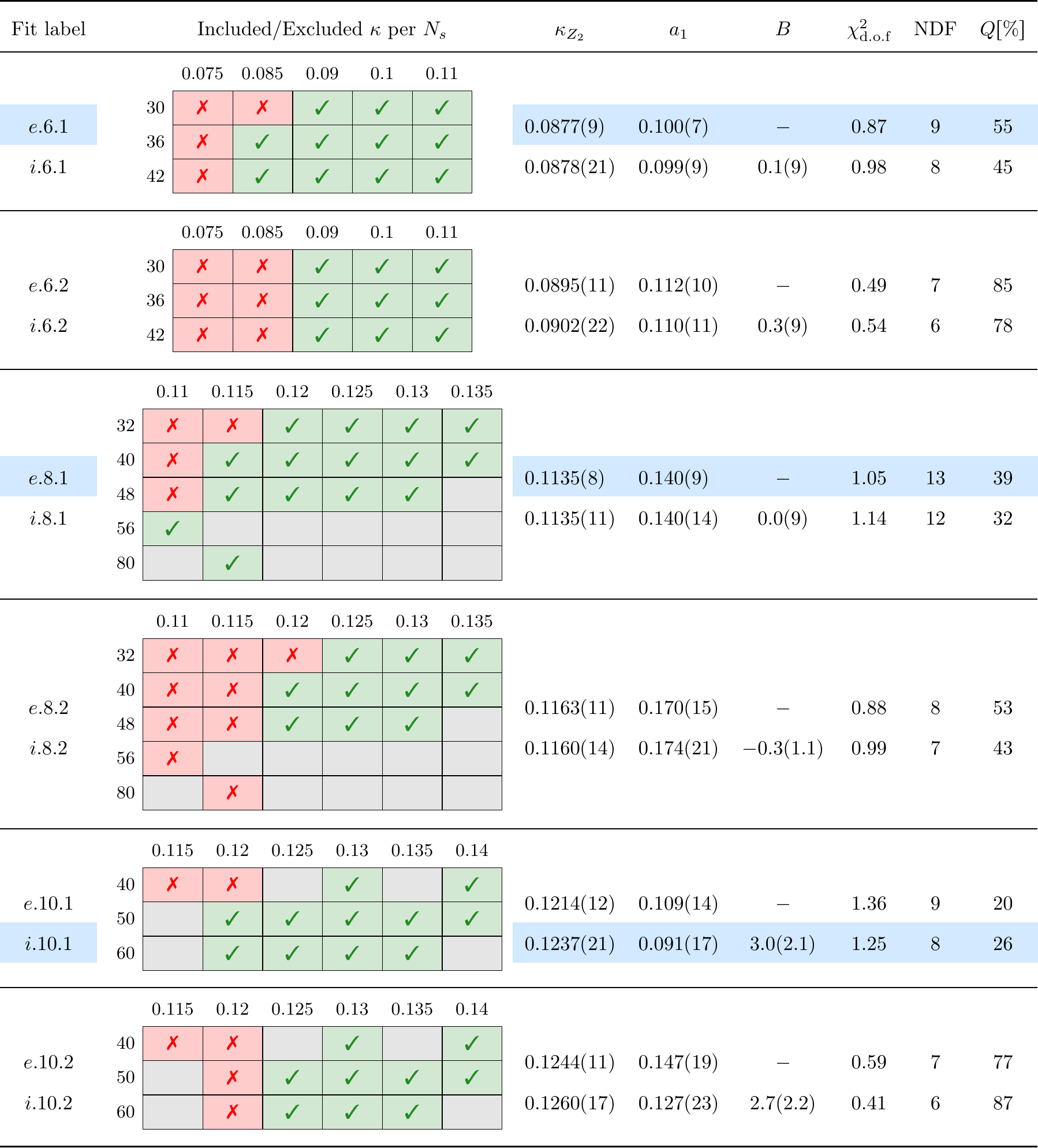}
    \caption{
      An overview of the outcome of the final fit analysis is presented here.
      For each value of $\Ntau$ the effect of excluding some data points is shown.
      The fits are labeled by $x.y.z$, where $x$ refers to the fit ansatz, $y$ is the value of $\Ntau$ and $z$ simply a counter.
      For $x=e$ the fit has been done according to \cref{eq:fitting_form_of_b4} (\emph{excluding} the correction term, $B="-"$), while for $x=i$ \cref{eq:kurtosis_finite_volume_correction_term} has been used and the correction term has been \emph{included}.
      $z=1$ always represents the fit with the least excluded data points, for $z>1$, more and more data points get excluded.     
      The rows with blue background contain the best fit as a compromise between all parameters.
      Subtables in the second column show which $\kappa$ (columns) have been included (\tc{ForestGreen}{\ding{51}}) or excluded (\tc{red}{\ding{55}}) for which $\Nspat$ (rows).
      Gray cells denote simulations that were not done.
     }
    \label{tab:mu0_fits_all_nt_with_kappa}
\end{table*}
\begin{figure*}[t]
    \centering
    \subfigure[Fit $e.6.1$ at $\Ntau=6$.]{ \label{fig:fitNt6}\includegraphics[width=0.98\columnwidth, clip, trim=4mm 4mm 0 0]{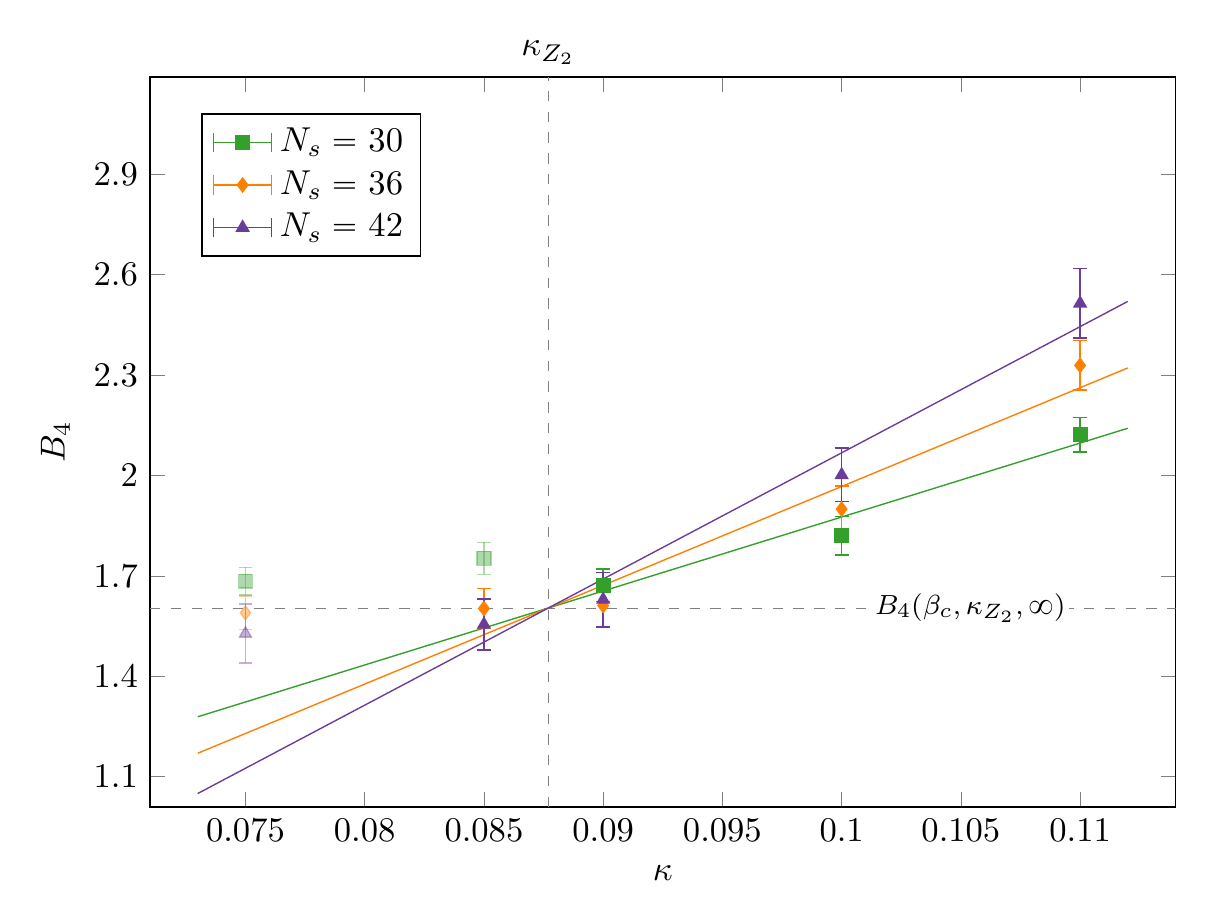}}\hfill
    \subfigure[Fit $e.8.1$ at $\Ntau=8$.]{ \label{fig:fitNt8}\includegraphics[width=0.98\columnwidth, clip, trim=4mm 4mm 0 0]{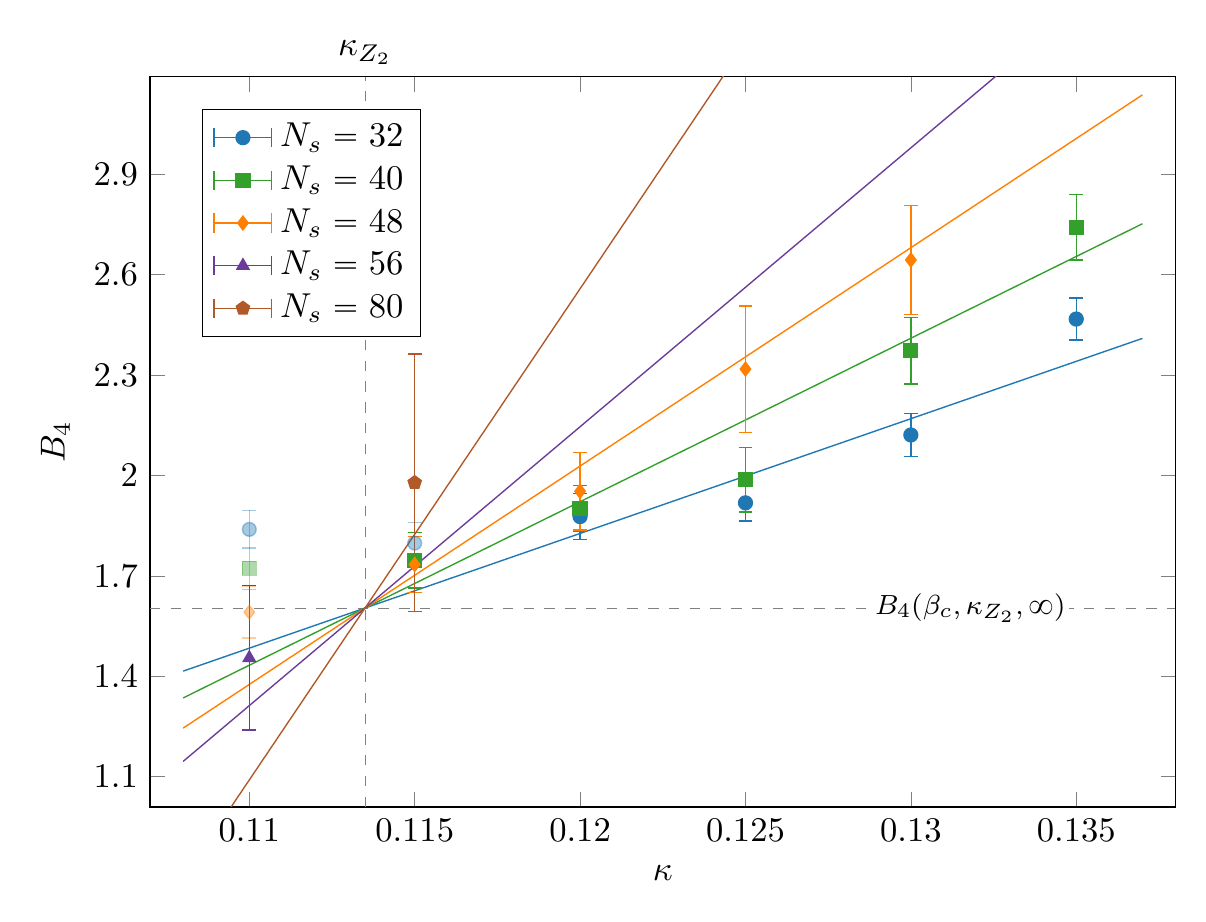}}\\
    \subfigure[Fit $i.10.1$ at $\Ntau=10$.]{\label{fig:fitNt10}\includegraphics[width=0.98\columnwidth, clip, trim=4mm 4mm 0 0]{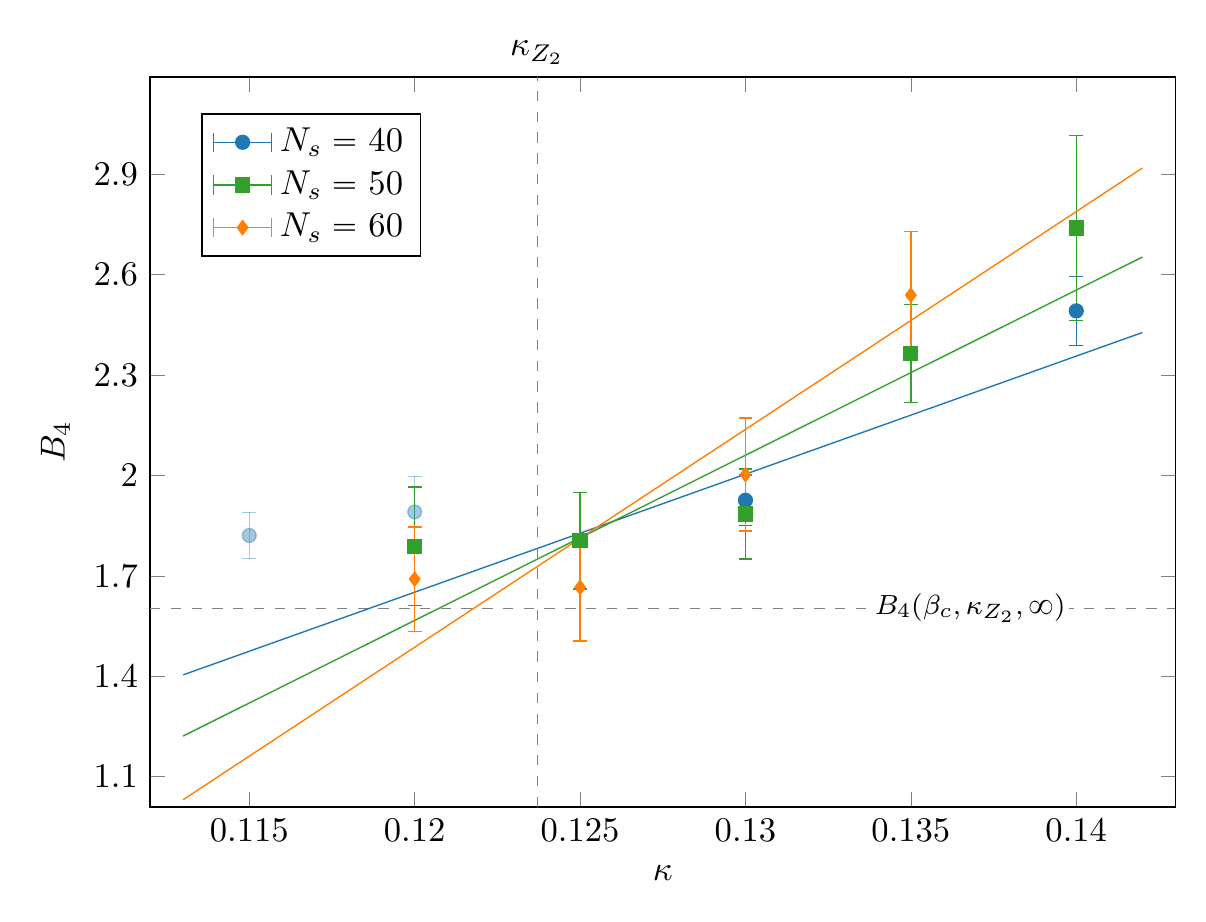}}
    \caption{%
      Final $\Kurtosis$ fits for $\Ntau\in\{6,8,10\}$.
      These refer to the coloured fields in \cref{tab:mu0_fits_all_nt_with_kappa}.
      Shaded points have not been included in the fits.
    }
    \label{fig:B4}
\end{figure*}
\begin{figure}[t]
    \centering
    \includegraphics[width=0.97\columnwidth, clip, trim=3mm 2mm 0 0]{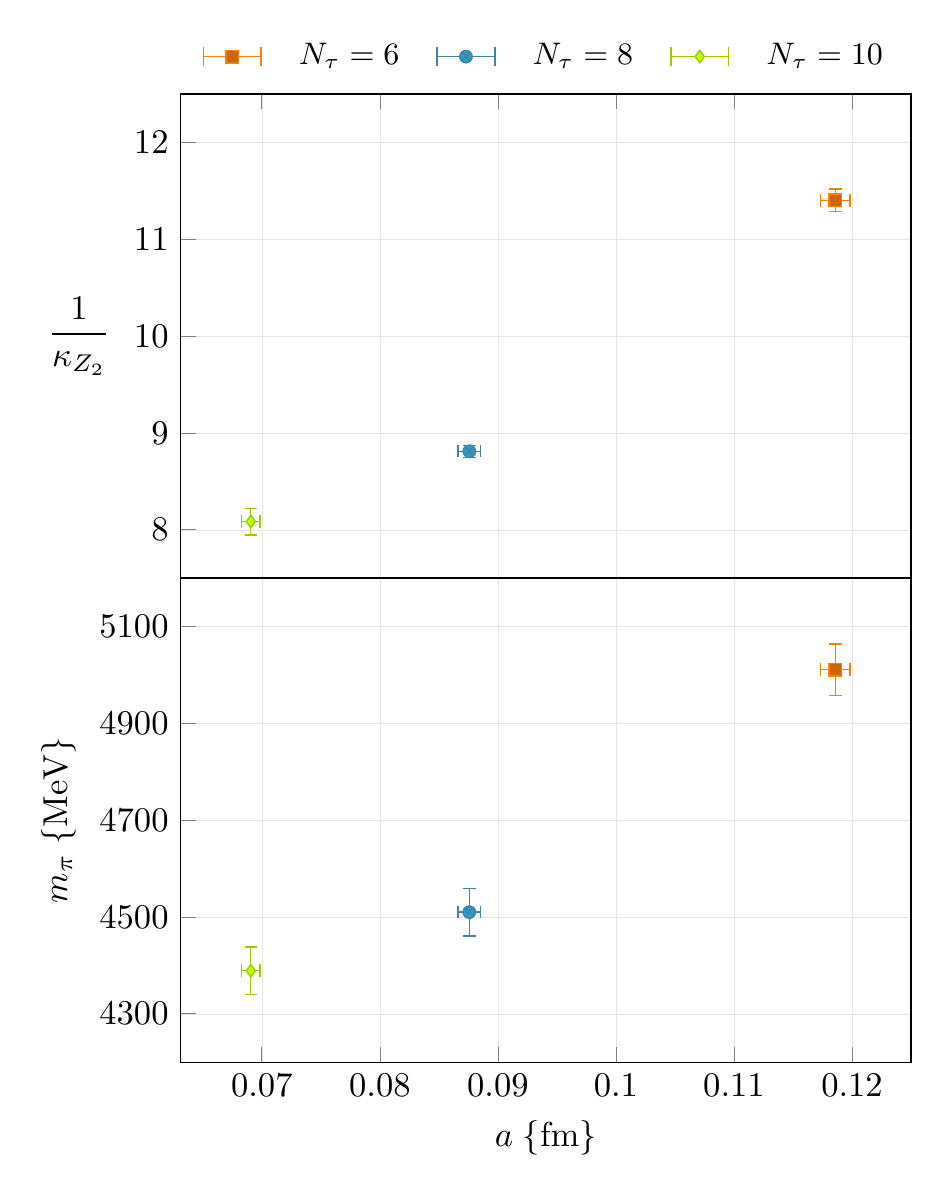}
    \caption{Above: The $\kc$ for three different lattice spacings. Below: Pion masses corresponding to the critical parameters.}
    \label{fig:result}
\end{figure}
We are now ready to extract our desired results from fits of our data to
\cref{eq:fitting_form_of_b4,eq:kurtosis_finite_volume_correction_term},
measuring the fit quality by the reduced chi-square $\chidof$ and
the $Q$-parameter. The latter amounts to the probability of getting a value of
chi-square larger than $\chidof$, assuming that the data have Gaussian noise,
and gives a measure of the quality of the fit (its optimal value is $50\%$).
In the ideal situation of negligible finite size effects, we expect to obtain
a good fit to \cref{eq:fitting_form_of_b4} and, simultaneously, a good fit to
\cref{eq:kurtosis_finite_volume_correction_term} with a consistent $\kc$ and
$B$ compatible with zero.
The general strategy then is to compare the fits performed with and without
the correction term described in \cref{eq:kurtosis_finite_volume_correction_term},
with the goal to isolate the leading terms.
For this purpose, it is in principle enough to
successively exclude points from the smallest volumes in the fit to
\cref{eq:kurtosis_finite_volume_correction_term}, until the coefficient $B$ is
compatible with zero and check for consistency of the fit parameters.

However, we generally observe that the inclusion of all volumes corresponding to the
smallest $\kappa$-values, i.e., the ones deepest in the  first-order region,
significantly deteriorates the fit quality, irrespective of the ansatz.
Because tunneling between the phases is exponentially suppressed by
volume in the first-order region, these points are increasingly difficult
to determine accurately, as discussed in the previous section.
Moreover, the entire $\Kurtosis$ curve is asymmetric about its inflection
point, because its two asymptotes are not symmetrically displaced
with respect to $\Kurtosis(\betaC,\kc,\infty)$.
As the entries in \cref{tab:nf2_nt6_nt8_nt10_mu0_data} illustrate,
the masses associated with the
smaller $\kappa$-values rise very quickly and are thus quite 
far from the critical quark mass in which we are interested.
Including these smallest $\kappa$-values
would therefore require higher order terms in the Taylor expansion of the kurtosis, 
in both brackets of \cref{eq:kurtosis_finite_volume_correction_term},
in order to account for the larger distance from the critical point as well as for the larger
finite size corrections. This would increase the number of fit parameters and weaken
the conclusions from the fits, so we excluded these points.
On the other hand, linear fits over ranges in $\kappa$ that only cover the crossover region
can result in biased estimates for $\kc$ (cf. the shift in $\kc$ comparing the
two $e.8.$\texttt{\#} and $i.8.$\texttt{\#} fits at $\Ntau=8$ represented in \cref{tab:mu0_fits_all_nt_with_kappa}).
Consequently, we made sure that for each $\Ntau$
we have included at least one $\kappa$ in the fits,
which reliably belongs to the first-order region. 

Distinctive features for $\kappa$ values in the first-order region are the
reversed $\Nspat$ ordering of the corresponding kurtosis (central) values with
respect to what happens in the crossover region (the value of the kurtosis
decreases with $\Nspat$ in the first-order region), as well as a more pronounced
two-peak structure in the distribution of the order parameter, which
we explicitly checked in every case. In fact, all data points with $\kappa<\kc$
fulfill these criteria, including the ones omitted from our
final fits.
For $\Ntau=8$ and $\kappa=0.115$, a parameter set in doubt, 
we simulated aspect ratio $10$ specifically
to check these features.
As can be seen in \cref{fig:fitNt8}, no 
hints for a first-order phase transition were found
which is consistent with the fit indicating $\kc<0.115$.
In our general strategy it is important that the fit gives a $\kc$ larger than the
smallest simulated $\kappa$ value, and that the $\kc$ extracted from the fit is 
cross-checked against
evidence for a first-order phase transition for all simulated $\kappa<\kc$.

Some representative fits are shown in \cref{tab:mu0_fits_all_nt_with_kappa}, where the
best ones chosen as our final result are highlighted.
Note that for $\Ntau=6,8$ we are able to find good fits with $B=0$, as well
as several consistent ones including additional data points and $B\neq 0$.
This indicates that all simulation data are indeed described by
\cref{eq:kurtosis_finite_volume_correction_term} and the coefficient $B$ is fully
controlled.
Our data for $\Kurtosis$ together with the best fits highlighted in the table
are also shown graphically in \cref{fig:B4}.

Our final results for the critical hopping parameter $\kc$ as a function of lattice
spacing are collected in \cref{fig:result} (top). Strong cut-off effects
are apparent. To compare with other approaches and get a feeling for the 
physical scales involved, \cref{fig:result} (bottom) shows the critical couplings 
converted to pseudoscalar meson masses. As already indicated in \cref{sec:simulation_details},
these numbers have to be taken with care because $a\mpi>1$ in all cases.
This problem could in principle be circumvented by heavy quark effective theory 
methods \cite{Sommer:2010ic}. However, it is apparent that at least two finer lattice
spacings are needed before a continuum extrapolation can be attempted. For these
the mass in lattice units might well be small enough, so we postpone this issue until
such data are available. Nevertheless, the shift in the critical pion mass is significantly
reduced between $\Ntau=8,10$ compared to $\Ntau=6,8$. A linear extrapolation using the 
last two points (unimproved Wilson fermions have ${\cal O}(a)$ effects) would then
predict $\mpi\approx 4$ GeV, where twice the shift to the extrapolated value should amount
to a conservative estimate of the remaining systematic error of $\sim 20\%$.

It is now interesting to compare our results with those of \refcite{Ejiri:2019csa},
where the critical points for $\Nf\in\{1,2,3\}$ Wilson fermions were determined 
on $\Ntau\in\{4,6\}$ by means of reweighting with a next-to-leading-order (NLO) hopping expanded
fermion determinant and a histogram technique instead of the 
cumulant analysis. For $\Nf=2$ on $\Ntau=6$, $\Nspat=24$, 
\refcite{Ejiri:2019csa} reports $\kc=0.1202(19)$ at NLO hopping expansion,
which translates to $\mpi^c/\Tc\approx 11.2$. 
Compared with our $\kc=0.0877(9)$ or $\mpi^c/\Tc\approx 18.1$ (which
is obtained from our scale setting summarized in \cref{tab:nf2_nt6_nt8_nt10_mu0_data}), one observes
a large discrepancy of $\approx$ 50\% in the critical pion mass. Note that,
since the same lattice action is employed, this discrepancy is not related
to the cut-off error on the determination of $\mpi^c$.
As discussed in \refcite{Ejiri:2019csa}, the histogram method is applied
at a fixed spatial volume and does not include an extrapolation to the 
thermodynamic limit. Indeed, the authors report a reduction of $\kc$ by 
$\approx 6\%$ when increasing volume to $\Nspat=32$ and by $\approx 13\%$ on $\Nspat=24$ 
when going from LO to 
NLO in the hopping expansion, i.e., both truncations have a systematic error
in the same direction. The comparison to our result highlights the necessity for either
refined approximations or a full calculation in order to achieve
a reliable continuum limit.

\section{Conclusions}\label{sec:conclusions}
While qualitatively well understood, the heavy mass (top right) corner of the Columbia 
plot in \cref{fig:discretization_effects_and_columbia_plot_schematic} (left) still lacks a 
quantitative determination, in the continuum limit, of the location of the deconfinement critical boundary.
In this work we focused our attention on the $\Nf=2$ deconfinement critical point and studied its location on progressively finer lattices simulating at $\Ntau\in \{6,8,10\}$.
Our results show that the continuum limit is not yet within reach given the extent of the observed cut-off effects. It is, indeed, apparent that for a continuum extrapolation to become feasible at least two finer lattice spacings will be needed.
Nevertheless, this work documents the progress that has been made both in refining the fitting strategy for the finite size scaling analysis and in appraising the growing statistics requirements towards the continuum limit.
	Concerning the former, we showed how a correction term can be used as a probe for finite size effects, which allowed us to isolate the leading terms in the linear kurtosis expansion around $\kc$ and identify the required aspect ratios for the linear regime.
With regards to the latter, our detailed analysis of the integrated autocorrelation times of the relevant observable illustrates the alarming prospects in terms of the statistics needed to reliably establish, as the continuum limit is approached, the order of the phase transition and the location of $\kc$.
Furthermore, our results allow for quantitative comparisons with those obtained using 
hopping expansions and thus
to assess their systematic error.

\begin{acknowledgments}
    We thank Christopher Czaban for collaboration during the early stages of this project
    \cite{Czaban:2016yae,Cuteri:2017zcb} and Reinhold Kaiser for providing a very handy GUI for the fitting analysis.
    The authors acknowledge support by the Deutsche Forschungsgemeinschaft (DFG, German Research Foundation) through the CRC-TR 211
    ``Strong-interaction matter under extreme conditions''~--~project number 315477589~--~TRR 211.
    We also thank the computing staff of the \hlr{} and \lcsc{} clusters for their support.
\end{acknowledgments}

\bibliographystyle{apsrev4-1}
\bibliography{./references}

\onecolumngrid

\appendix*
\section{Tables with scale setting, statistics and integrated autocorrelation time}

\begin{table}[!h]
    \setlength{\tabcolsep}{5mm}
    \renewcommand{\arraystretch}{1.1}
    \centering
    \newcommand{\CC}[1]{\cellcolor{#1}}
    \begin{tabular}{c*{2}{S[table-format=1.4]}S[table-format=1.4(1)]S[table-format=1.4(2)]S[table-format=2.3(2)]S[table-format=3(1)]}
        \toprule
        $\Ntau$ & {$\kappa$} & {$\betaC$} & {$a\,\mpi$} & $a\,\{\si{\femto\meter}\}$ & {$\mpi\,\{\si{\giga\electronvolt}\}$} & {$\Tc\,\{\si{\mega\electronvolt}\}$} \\ 
        \midrule
        & 0.075  & 5.8893 & 3.4722(2) & 0.1181(12) & 5.80(6) & 279(3) \\
        & 0.085  & 5.8845 & 3.1073(2) & 0.1185(12) & 5.17(5) & 277(3) \\
        \rowcolor{gray!30}[15pt][15pt]\CC{white}
        & 0.0877 & 5.8821 & 3.0111(2) & 0.1186(13) & 5.01(5) & 277(3) \\
        & 0.09   & 5.8798 & 2.9306(2) & 0.1191(13) & 4.86(5) & 276(3) \\
        & 0.1    & 5.8676 & 2.5810(2) & 0.1203(13) & 4.24(4) & 273(3) \\
        \multirow{-6}{*}{6}
        & 0.11   & 5.8462 & 2.2383(2) & 0.1232(14) & 3.58(4) & 267(3) \\
        \midrule
        & 0.11   & 6.0306 & 2.1298(2) & 0.0872(9)  & 4.82(5) & 283(3) \\
        \rowcolor{gray!30}[15pt][15pt]\CC{white}
        & 0.1135 & 6.0222 & 2.0017(2) & 0.0876(9)  & 4.51(5) & 282(3)  \\
        & 0.115  & 6.0180 & 1.9471(2) & 0.0887(9)  & 4.33(5) & 278(3) \\
        & 0.12  & 6.0009 & 1.7645(2) & 0.0893(10) & 3.90(4) & 276(3) \\
        & 0.125  & 5.9776 & 1.5814(2) & 0.0906(10) & 3.44(4) & 272(3) \\
        & 0.13   & 5.9464 & 1.3996(3) & 0.0926(10) & 2.98(3) & 266(3) \\
        \multirow{-7}{*}{8}
        & 0.135  & 5.9026 & 1.2212(3) & 0.0953(10) & 2.53(3) & 259(3) \\
        \midrule
        & 0.115  & 6.1682 & 1.8724(2) & 0.0680(8)  & 5.43(6) & 290(3) \\
        & 0.12   & 6.1543 & 1.6802(2) & 0.0688(8)  & 4.82(5) & 287(3) \\
        \rowcolor{gray!30}[15pt][15pt]\CC{white}
        & 0.1237 & 6.1414 & 1.5361(2) & 0.0691(8)  & 4.39(5) & 286(3) \\
        & 0.125  & 6.1356 & 1.4858(2) & 0.0694(8)  & 4.23(5) & 284(3) \\
        & 0.13   & 6.1027 & 1.2930(2) & 0.0712(8)  & 3.58(4) & 277(3) \\
        & 0.135  & 6.0576 & 1.0999(4) & 0.0720(8)  & 3.01(3) & 274(3) \\
        \multirow{-7}{*}{10}
        & 0.14   & 5.9902 & 0.9143(4) & 0.0761(8)  & 2.37(3) & 259(3) \\
        \bottomrule
    \end{tabular}
    \caption{%
      Outcome of the pion mass and the scale setting simulations, performed on $32 \times 16^3$ lattices, accumulating 800 independent configurations.
      The values of $\betaC$ and the critical temperature $\Tc$ of the deconfinment phase transition are also included to provide a more complete overview.
      $a$ and $\mpi$ have also been measured at the $\kc$-values obtained from the fits and are reported on gray background.
      The value of $\betaC$ at $\kc$, at which the scale is set and the pion mass is 
	measured, has been obtained via interpolation of $\betaC$-values at simulated $\kappa$ nearby.
    }
    \label{tab:nf2_nt6_nt8_nt10_mu0_data}
    \vspace{10mm}
    \renewcommand{\arraystretch}{1.1}
    \setlength{\tabcolsep}{5mm}
    \centering
    \begin{tabular}{cS[table-format=1.3]*{4}{c}}
        \toprule
        \multirow{2}{*}{$\Ntau$} & {\multirow{2}{*}{$\kappa$}} &  \multicolumn{4}{c}{$\betaC$  | Total statistics per $N_{s}$  | Number of simulated $\beta$ values }\\
        &         &        Aspect ratio $4$ & Aspect ratio $5$ & Aspect ratio $6$ & Aspect ratio $7$ \\
        \midrule
        \multirow{5}{*}{6}
        & 0.075 & -- & 5.88884 | 1.6M | 2 & 5.88895 | 1.6M | 2  & 5.88933 | 1.6M | 2 \\
        & 0.085 & -- & 5.88407 | 1.6M | 2 & 5.88448 | 1.6M | 2  & 5.88452 | 1.6M | 2 \\
        & 0.09  & -- & 5.88097 | 2.4M | 3 & 5.88104 | 2.4M | 3  & 5.87985 | 2.4M | 3 \\
        & 0.1   & -- & 5.86865 | 1.6M | 2 & 5.86762 | 1.6M | 2  & 5.86758 | 1.6M | 2 \\
        & 0.11  & -- & 5.84677 | 1.6M | 2 & 5.84624 | 2.4M | 3  & 5.84623 | 2.4M | 3 \\
        \midrule
        \multirow{6}{*}{8}
        & 0.11  & 6.03018 | 2.4M | 3 & 6.03085 | 2.4M | 3 & 6.03064 | 2.4M | 3 & 6.03139 | 1.0M | 3 \\
        & 0.115 & 6.01892 | 2.4M | 3 & 6.01891 | 2.4M | 3 & 6.01801 | 2.4M | 3 & -- \\
        & 0.12  & 6.00366 | 1.6M | 2 & 6.00208 | 2.4M | 3 & 6.00093 | 0.9M | 2 & -- \\
        & 0.125 & 5.98070 | 2.4M | 3 & 5.97888 | 2.2M | 3 & 5.97757 | 1.4M | 3 & -- \\
        & 0.13  & 5.94928 | 2.4M | 3 & 5.94705 | 2.4M | 3 & 5.94642 | 1.6M | 2 & -- \\
        & 0.135 & 5.90492 | 2.4M | 3 & 5.90257 | 2.4M | 3 & -- & -- \\
        \midrule
        \multirow{6}{*}{10}
        & 0.115 & 6.16818 | 1.8M | 3 &  -- & -- & -- \\
        & 0.12  & 6.15297 | 1.8M | 3 & 6.15408 | 1.8M | 3 & 6.15434 | 1.6M | 3 & -- \\
        & 0.125 & --                 & 6.13558 | 1.8M | 3 & 6.13558 | 1.2M | 2 & -- \\
        & 0.13  & 6.10685 | 1.8M | 3 & 6.10524 | 1.8M | 3 & 6.10269 | 1.2M | 2 & -- \\
        & 0.135 & --                 & 6.05851 | 1.8M | 3 & 6.05758 | 1.6M | 4 & -- \\
        & 0.14  & 5.99361 | 1.8M | 3 & 5.99022 | 1.8M | 3 & -- & -- \\
        \bottomrule
    \end{tabular}
    \captionof{table}{%
      Statistics overview of $\Ntau\in \{6,8,10\}$.
      For $\Ntau=8$, $\Nspat=80$ we simulated at three different $\beta$ values, $\betaC=6.01708$ , and have an overall statistics of 2.0M.
    }
    \label{tab:simulation_overview_nf2_mu_zero}
\end{table}

\begin{table}[ht]
    \renewcommand{\arraystretch}{1.2}
    \setlength{\tabcolsep}{4mm}
    \centering
    \begin{tabular}{cS[table-format=1.3]*{4}{S[table-format=2.6(3), parse-numbers=false]@{\hspace{1ex}|\hspace{1ex}}r}}
        \toprule
        \multirow{2}{*}{$\Ntau$} & {\multirow{2}{*}{$\kappa$}} &  \multicolumn{8}{c}{Average $\TauInt(\Skewness)\cdot10^{-2}$ | Average number of independent events }\\ 
        &         &        \multicolumn{2}{c}{Aspect ratio $4$} & \multicolumn{2}{c}{Aspect ratio $5$} & \multicolumn{2}{c}{Aspect ratio $6$} & \multicolumn{2}{c}{Aspect ratio $7$} \\
        \midrule
        \multirow{5}{*}{6}
        & 0.075 & \multicolumn{2}{c}{--} & 9.8(9)     & 437  & 18.2(2.0)  & 222  & 50(9)      & 95  \\
        & 0.085 & \multicolumn{2}{c}{--} & 10.6(1.0)  & 397  & 24(3)      & 187  & 39(6)      & 104 \\
        & 0.09  & \multicolumn{2}{c}{--} & 14.7(1.7)  & 381  & 24(3)      & 243  & 15.7(1.9)  & 351   \\
        & 0.1   & \multicolumn{2}{c}{--} & 9.6(9)     & 437  & 11.3(1.1)  & 382  & 10.1(1.0)  & 432   \\
        & 0.11  & \multicolumn{2}{c}{--} & 5.1(4)     & 790  & 5.07(29)   & 825  & 4.92(28)   & 864  \\
        \midrule
        \multirow{6}{*}{8}
        & 0.11  & 10.0(8)    & 419  & 17.5(1.6)  & 239  & 32(4)      & 139        & 37(6)  & 45            \\
        & 0.115 & 12.1(1.0)  & 343  & 17.9(1.7)  & 236  & 31(4)      & 146        & \multicolumn{2}{c}{--} \\
        & 0.12  & 9.6(8)     & 429  & 16.2(1.6)  & 271  & 15.2(2.2)  & 158        & \multicolumn{2}{c}{--} \\
        & 0.125 & 7.5(5)     & 550  & 8.9(7)     & 468  & 12.2(1.2)  & 202        & \multicolumn{2}{c}{--} \\
        & 0.13  & 5.7(3)     & 731  & 6.1(4)     & 683  & 5.9(4)     & 677        & \multicolumn{2}{c}{--} \\
        & 0.135 & 3.08(14)   & 1293 & 3.14(14)   & 1273 &  \multicolumn{2}{c}{--} & \multicolumn{2}{c}{--} \\
        \midrule
        \multirow{6}{*}{10}
        & 0.115 & 12.1(1.1)  & 258       &  \multicolumn{2}{c}{--} & \multicolumn{2}{c}{--} & \multicolumn{2}{c}{--} \\
        & 0.12  & 13.5(1.3)  & 233       &  30(4)      & 109       &  51(9)      & 65       & \multicolumn{2}{c}{--} \\
        & 0.125 & \multicolumn{2}{c}{--} &  20.9(2.3)  & 147       &  37(7)      & 87       & \multicolumn{2}{c}{--} \\
        & 0.13  & 13.5(1.3)  & 230       &  21.4(2.4)  & 147       &  21.1(2.8)  & 138      & \multicolumn{2}{c}{--} \\
        & 0.135 & \multicolumn{2}{c}{--} &  8.2(6)     & 379       &  8.9(7)     & 213      & \multicolumn{2}{c}{--} \\
        & 0.14  & 4.07(24)   & 765       &  3.85(22)   & 820       & \multicolumn{2}{c}{--} & \multicolumn{2}{c}{--} \\
        \bottomrule
    \end{tabular}
    \caption{%
      Average of the integrated autocorrelation time $\TauInt$ and of the number of independent events for the skewness of the order parameter.
      The average has been done among the merged chains at the simulated $\beta$.
    }
    \label{tab:tauint_B3}
    \vspace{15mm}
    \renewcommand{\arraystretch}{1.2}
    \setlength{\tabcolsep}{4mm}
    \centering
    \begin{tabular}{cS[table-format=1.3]*{4}{S[table-format=2.6(3), parse-numbers=false]@{\hspace{1ex}|\hspace{1ex}}r}}
        \toprule
        \multirow{2}{*}{$\Ntau$} & {\multirow{2}{*}{$\kappa$}} &  \multicolumn{8}{c}{Average $\TauInt(\Kurtosis)\cdot10^{-2}$ | Average number of independent events }\\ 
        &         &        \multicolumn{2}{c}{Aspect ratio $4$} & \multicolumn{2}{c}{Aspect ratio $5$} & \multicolumn{2}{c}{Aspect ratio $6$} & \multicolumn{2}{c}{Aspect ratio $7$} \\
        \midrule
        \multirow{5}{*}{6}
        & 0.075 & \multicolumn{2}{c}{--} &  3.07(17)  &  1313    & 13.7(1.4) &  292     &     31(4)  &  131  \\
        & 0.085 & \multicolumn{2}{c}{--} &  3.42(22)  &  1293    & 15.3(1.9) &  323     &     30(4)  &  139  \\
        & 0.09  & \multicolumn{2}{c}{--} & 10.8(1.4)  &   792    &   20(3)   &  308     &    8.1(5)  &  499  \\
        & 0.1   & \multicolumn{2}{c}{--} &  4.7(3)  &   857    &  8.3(7)   &  540     &    8.5(8)  &  524  \\
        & 0.11  & \multicolumn{2}{c}{--} &  3.18(20)  &  1399    & 3.47(17)  & 1216     &   3.41(17) & 1267  \\
        \midrule
        \multirow{6}{*}{8}
        & 0.11  & 5.0(4)   & 1049     &  9.5(8)   &  507      &     20.7(2.7) &  294    & 31(5) & 53 \\
        & 0.115 & 6.4(5)   &  845     & 12.8(1.2) &  362      &     17.2(2.0) &  331    & \multicolumn{2}{c}{--} \\
        & 0.12  & 3.59(20) & 1105     &  7.8(6)   &  592      &      8.1(9)   &  281    & \multicolumn{2}{c}{--} \\
        & 0.125 & 3.86(21) & 1116     &  4.96(29) &  739      &      9.4(1.0) &  288    & \multicolumn{2}{c}{--} \\
        & 0.13  & 3.12(17) & 1508     &  3.81(20) & 1128      &      3.61(21) & 1102    & \multicolumn{2}{c}{--} \\
        & 0.135 & 1.74(6)  & 2381     &  2.05(8)  & 1945      &  \multicolumn{2}{c}{--} & \multicolumn{2}{c}{--} \\
        \midrule
        \multirow{6}{*}{10}
        & 0.115 &    3.41(18) & 892      &  \multicolumn{2}{c}{--} & \multicolumn{2}{c}{--} & \multicolumn{2}{c}{--} \\
        & 0.12  &    5.6(4)   & 576      &     21(3)     &  215    &     28(4)     & 103    & \multicolumn{2}{c}{--} \\
        & 0.125 & \multicolumn{2}{c}{--} &     12.0(1.3) &  307    &     21(3)     & 165    & \multicolumn{2}{c}{--} \\
        & 0.13  &    4.7(3)   & 722      &     13.4(1.5) &  261    &     11.9(1.3) & 244    & \multicolumn{2}{c}{--} \\
        & 0.135 & \multicolumn{2}{c}{--} &      5.0(3)   &  596    &      5.7(4)   & 336    & \multicolumn{2}{c}{--} \\
        & 0.14  &    2.20(10) & 1393     &      2.17(10) & 1429    &      \multicolumn{2}{c}{--}    & \multicolumn{2}{c}{--} \\
        \bottomrule
    \end{tabular}
    \captionof{table}{%
      Average of the integrated autocorrelation time $\TauInt$ and of the number of independent events for the kurtosis of the order parameter.
      The average has been done among the merged chains at the simulated $\beta$.
    }
    \label{tab:tauint_B4}
\end{table}

\end{document}